\documentclass[11pt]{article}

\usepackage{array, color, enumerate, enumitem, epstopdf, epsfig, epsf,
  float, geometry, graphicx, helvet, ifthen, latexsym, lscape,
  multicol, overpic, setspace, subfigure, titlesec, ulem,
  url, etoolbox, natbib, bm}

\usepackage{amsbsy, amsfonts, amsmath, amssymb, amsthm}

\usepackage[colorlinks,linkcolor=blue,citecolor=blue,urlcolor=blue,filecolor=blue]{hyperref}

\newcommand{\real}{\mathbb{R}} 
\newcommand{\simp}{\mathbb{S}} 

\def\bx{\boldsymbol{x}}

\def\x{\boldsymbol{x}}
\def\y{\boldsymbol{y}}
\def\R{\boldsymbol{R}}
\def\r{\boldsymbol{r}}
\def\X{\boldsymbol{X}}

\def\bD{\boldsymbol{D}}

\newcommand{\be}{\begin{equation}}
\newcommand{\ee}{\end{equation}}
\newcommand{\bea}{\begin{eqnarray}}
\newcommand{\eea}{\end{eqnarray}}
\newcommand{\bes}{\begin{eqnarray*}}
\newcommand{\ees}{\end{eqnarray*}}
\newcommand{\bi}{\begin{itemize}}
\newcommand{\ei}{\end{itemize}}

\def\tx{\widetilde \bx}

\usepackage{lineno}

\newcommand*\linenomathpatch[1]{%
  \cspreto{#1}{\linenomath}%
  \cspreto{#1*}{\linenomath}%
  \csappto{end#1}{\endlinenomath}%
  \csappto{end#1*}{\endlinenomath}%
}
\newcommand*\linenomathpatchAMS[1]{%
  \cspreto{#1}{\linenomathAMS}%
  \cspreto{#1*}{\linenomathAMS}%
  \csappto{end#1}{\endlinenomath}%
  \csappto{end#1*}{\endlinenomath}%
}

\expandafter\ifx\linenomath\linenomathWithnumbers
  \let\linenomathAMS\linenomathWithnumbers
  \patchcmd\linenomathAMS{\advance\postdisplaypenalty\linenopenalty}{}{}{}
\else
  \let\linenomathAMS\linenomathNonumbers
\fi

\linenomathpatch{equation}
\linenomathpatchAMS{gather}
\linenomathpatchAMS{multline}
\linenomathpatchAMS{align}
\linenomathpatchAMS{alignat}
\linenomathpatchAMS{flalign}

\makeatletter
\patchcmd{\mmeasure@}{\measuring@true}{
  \measuring@true
  \ifnum-\linenopenaltypar>\interdisplaylinepenalty
    \advance\interdisplaylinepenalty-\linenopenalty
  \fi
  }{}{}
\makeatother

\usepackage{algorithm,algpseudocode}
\newtheorem{definition}{Definition}
\newlength\algoindent
\renewcommand\algorithmicrequire{\textbf{Parameters:}}

\usepackage{wrapfig}
\usepackage{pgfplots}
\usepackage{tikz}

\graphicspath{{./}}

\newcommand*{\email}[1]{%
    \href{mailto:#1}{\color{black}{#1}}\par
    }

\begin{document}

\title{Principal Amalgamation Analysis for Microbiome Data}
\author{{Yan Li$^1$},
  {Gen Li$^2$},
  {and Kun Chen$^{1, }$}\thanks{Corresponding author. Email: \email{kun.chen@uconn.edu}}\\\\
  $^1$Department of Statistics, University of Connecticut\\
  $^2$Department of Biostatistics, University of Michigan, Ann Arbor
}

\maketitle

\begin{abstract}
  In recent years microbiome studies have become increasingly
  prevalent and large-scale. Through high-throughput sequencing
  technologies and well-established analytical pipelines, relative
  abundance data of operational taxonomic units and their associated
  taxonomic structures are routinely produced. Since such data can be
  extremely sparse and high dimensional, there is often a genuine need
  for dimension reduction to facilitate data visualization and
  downstream statistical analysis. We propose \textit{Principal
    Amalgamation Analysis} (PAA), a novel amalgamation-based and
  taxonomy-guided dimension reduction paradigm for microbiome
  data. Our approach aims to aggregate the compositions into a smaller
  number of \textit{principal compositions}, guided by the available
  taxonomic structure, by minimizing a properly measured loss of
  information. The choice of the loss function is flexible and can be
  based on familiar diversity indices for preserving either
  within-sample or between-sample diversity in the data. To enable
  scalable computation, we develop a hierarchical PAA algorithm to
  trace the entire trajectory of successive simple
  amalgamations. Visualization tools including dendrogram, scree plot,
  and ordination plot are developed. The effectiveness of PAA is
  demonstrated using gut microbiome data from a preterm infant study
  and an HIV infection study.
\end{abstract}

\vfill

\section{Introduction}

Microbiome, the set of microorganisms inhabiting a specific biological niche, plays a critical role in the development, nutrition, immunity, and health of their host organisms \citep{turnbaugh2007human}. 
The human gut microbiome, for instance, is known to not only control digestion but also affect immune and nervous systems of human \citep{turnbaugh2006obesity,tremlett2017gut,kau2011human}. Either being host associated or free living, microbial communities are indispensable components of their ecosystems, and a deep understanding of their community structure and their interactions with environment could lead to important biological and ecological insights. Indeed, in recent years, microbiome studies have become increasingly prevalent and large-scale, also owing to the rapid advances in high-throughput sequencing technologies. The raw sequencing reads can now be processed by well-established analytical pipelines such as quantitative insights into microbial ecology (QIIME) and mothur, to produce abundance tables of operational taxonomic units (OTU) \citep{Schloss2009,QIIME2010,Chong2020}. Since the number of sequencing reads per sample (i.e., library size) may vary dramatically, proper sampling and normalization procedures are further adopted to produce relative abundance data of the OTUs \citep{gloor2016s,tsilimigras2016compositional}, from which downstream analysis is performed.

Microbiome data is complex in nature, subject to constraints such as compositionality, high dimensionality, zero inflation, over dispersion, and taxonomic hierarchy. Specifically, microbiome data, as presented in relative abundances or proportions, are compositional; each compositional vector resides in a simplex that does not admit the standard Euclidean geometry. Second, the date are often very sparse with a large portion of zeros, arising from either under-sampling or true absence of the corresponding taxa. Third, the number of OTUs or taxon is often much larger than the number of samples, making the data analysis prone to many curses of dimensionality. In addition, a unique feature of microbiome data is the presence of the evolutionary history of the taxa charted through a taxonomic tree. This hierarchical structure provides crucial information about the relationship between different microbes and is proven useful in various studies \citep{randolph2018kernel,xiao2018phylogeny,tanaseichuk2013phylogeny,garcia2013identification}. These inherent characteristics of microbiome data impose various statistical challenges and stress the need for developing novel methods to better harness the power of such data.

As the microbiome data is often extremely sparse and high dimensional in many studies, there is a genuine need to properly reduce its dimension to facilitate data exploration, visualization, and downstream analysis. Besides some ``naive'' data reduction methods such as directly using certain diversity measures as coarse summaries or keeping only the most prevalent taxon, ordination methods such as principal component analysis (PCA) and principal coordinate analysis (PCoA) are among the most commonly-adopted approaches in practice. PCA generally relies on transformations that neglect the unique features of microbiome data \citep{aitchison1983principal,aitchison2002biplots}, such as zero-inflation and taxonomic tree structure; PCoA, on the other hand, is based on a proximity matrix that may accommodate the data features but fails to pinpoint relevant microbes that drive the data reduction \citep{lozupone2011unifrac}. Moreover, for microbiome data with taxonomic information, there is a trade-off between data resolution and accuracy: the lower the taxonomic rank, the higher the data resolution (with more taxa and thus more information), but the sparser and less accurate the data (there are more zeros and each composition is converted from a smaller count). Most existing methods only apply to a prefixed taxonomic rank and/or rely on transformations (with ad-hoc replacement of zeros) that may inflate inaccuracy \citep{Weiss2017}. These often lead to unstable and biased results \citep{Palarea2013,mcmurdie2014waste}.

We concern a fundamental question: {\it what constitutes an interpretable and effective dimension reduction of microbiome data?} It is apparent that the answers from the aforementioned approaches are with flaws. Our answer is radically different and yet strikingly intuitive: we argue that for microbiome (relative abundance) data, an effective and interpretable operation for dimension reduction is through aggregating the compositional components, i.e., through the so-called \textit{amalgamation}, a fundamental operation on compositional data \citep{aitchison1982statistical}. 
More precisely, if the components of a length-$p$ compositional vector are separated into $k < p$ mutually exclusive and exhaustive subsets and the components of each subset are added together, the resulting length-$k$ compositional vector is termed an amalgamation. For instance, $(x_1 + x_2, x_3, x_4 +x_5)$ is an amalgamation of $(x_1, x_2, x_3, x_4, x_5)$. Given its simplicity, it is not surprising that amalgamation has been widely used in practice, but mostly in a rather ad-hoc way, e.g., combining a number of compositional components with the lowest prevalence. Not until recently, a few studies on amalgamation-based dimension reduction emerged \citep{Greenacre2021Amalgam,Quinn2020Amalgam}. A more detailed review on exisiting dimension reducton methods for microbiome data is provide in Supplementary Information \ref{sec:supp:literature}.  

We propose \textit{Principal Amalgamation Analysis} (PAA), an amalgamation-based and taxonomy-guided dimension reduction paradigm for microbiome data. Our PAA approach directly handles the compositional data without the need of transformation and reduces its dimension by clustering and aggregating the compositions based on minimizing certain information loss, subject to confinement to the taxonomic hierarchy. The choice of the loss function can be flexible and problem specific; for example, it can be based on diversity measures such as $\alpha$ diversity and $\beta$ diversity to examine and preserve either within-sample (between-species) or between-sample (between-habitat) diversity of the data. To enable scalable computation, we develop and implement an efficient agglomerative clustering algorithm to identify the entire trajectory of the successive simple amalgamation steps. This allows us to start from the raw OTUs at their lowest taxonomic ranks and gradually amalgamate them until a desired balance between information loss and dimension reduction is reached. As such, PAA alleviates the bias and instability introduced by zero replacement and data transformations, maintains the compositional and taxonomic structures of the reduced data, and offers superior interpretation and visualization through the resulting ``principal compositions''.



\section{Setup with An Illustrative Example}\label{sec:setup}

To illustrate the proposed taxonomy-guided dimension reduction, we consider a preterm infant study conducted at a Neonatal Intensive Care Unit (NICU) in the northeast region of the U.S. Fecal samples of preterm infants were collected daily when available during the infant’s first month of postnatal age. Bacterial DNA was isolated and extracted from each sample \citep{Bomar2011,cong2017influence}; V4 regions of the 16S rRNA gene were sequenced using the Illumina platform and clustered and analyzed using QIIME \citep{QIIME2010} to produce microbiome data. 
When infant reached 36-38 weeks of postmenstrual age, neurobehavioral outcomes were measured using the NICU Network Neurobehavioral Scale (NNNS) \citep{cong2017influence}. The main interest was about examining the gut-brain axis, i.e., whether and how gut microbiome compositions during early postnatal stage impact later neurobehavioral outcomes.

\begin{figure*}[!h]
\centering
\subfigure[Heatmap]{\includegraphics[width=0.46\textwidth]{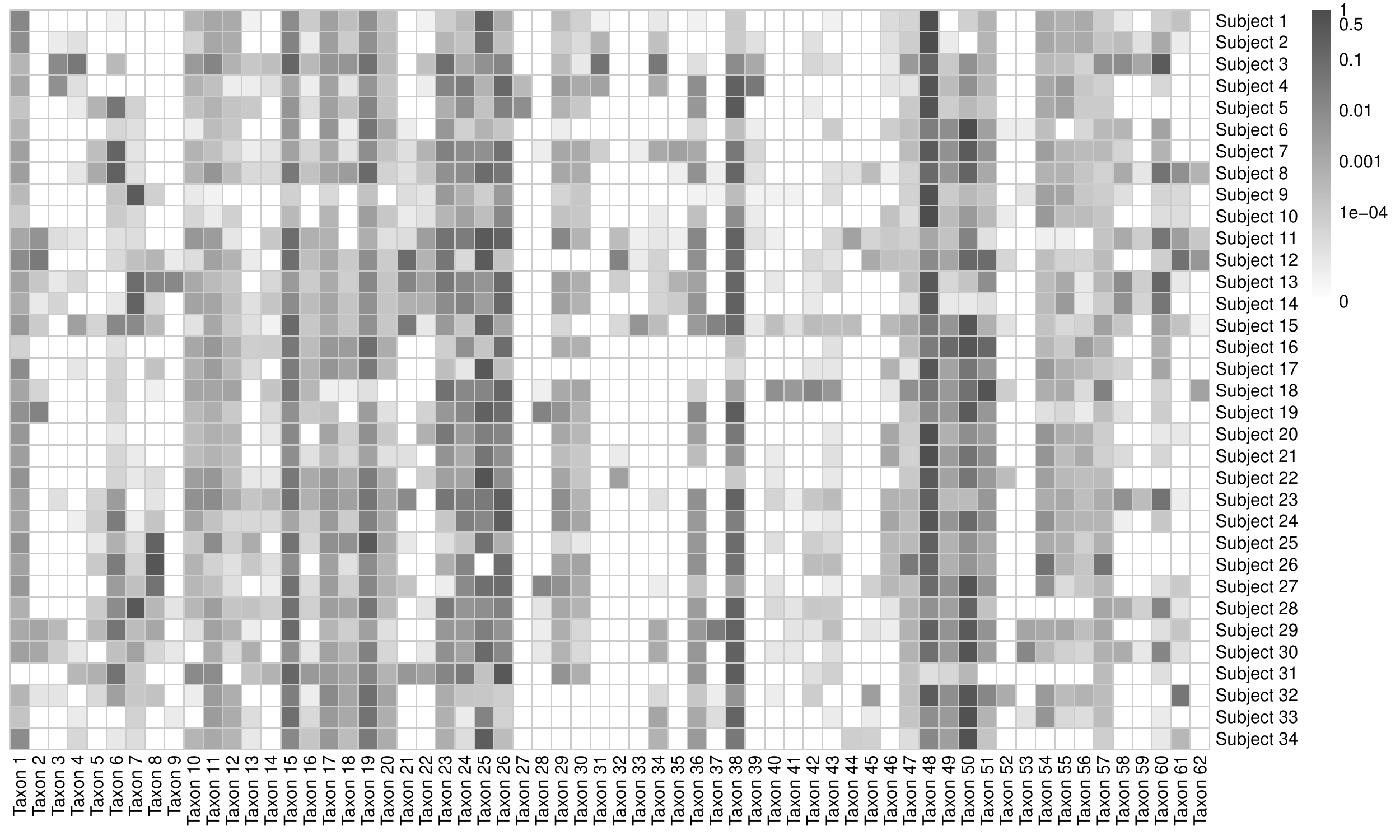}}
\subfigure[Barplot of relative abundance]{\includegraphics[width=0.46\textwidth]{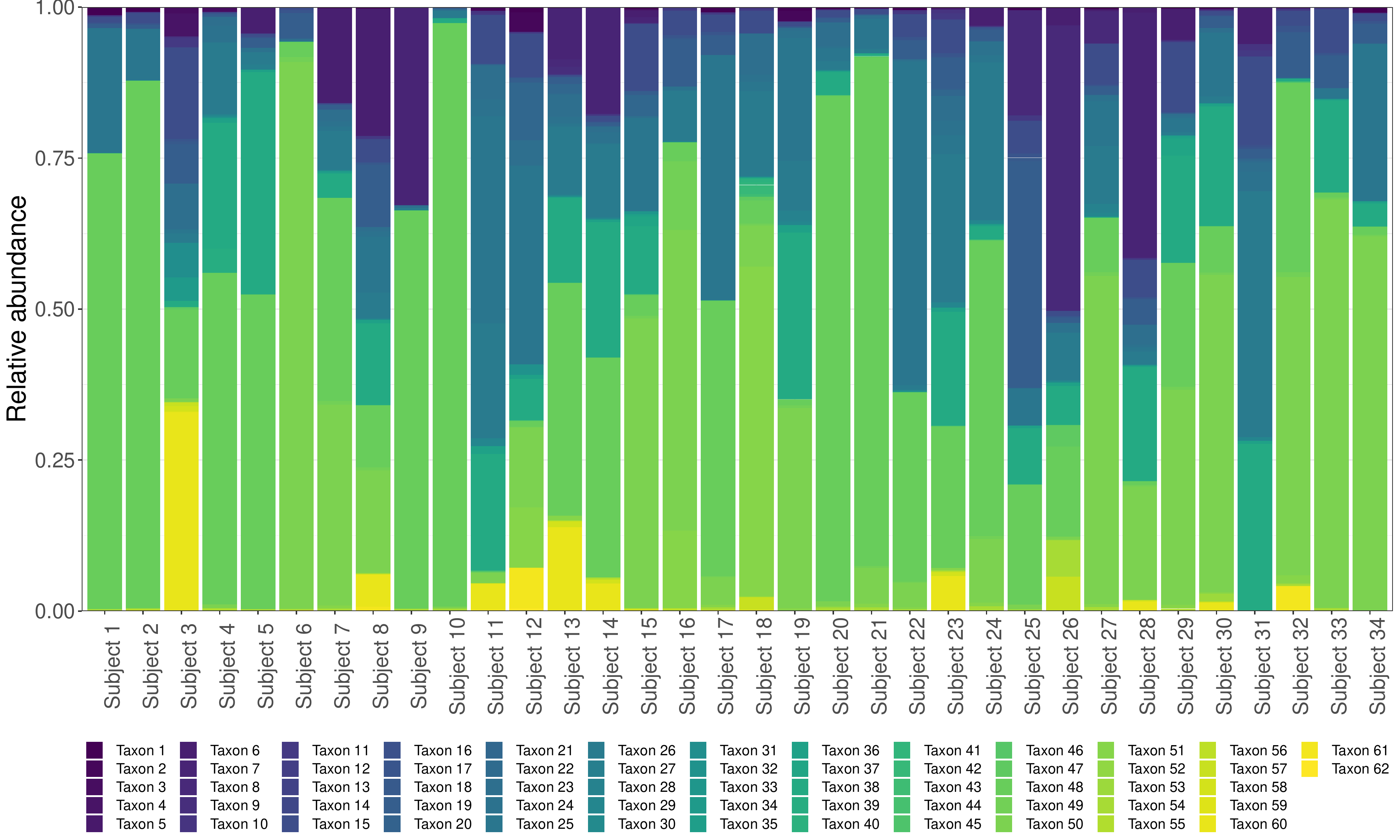}}     
 \caption{{\footnotesize The NICU data: Heatmap and barplot of the relative abundance data}}\label{fig:taxonomy_nicu}
\end{figure*}

\begin{figure*}[htp]
\centering
       \includegraphics[width=11.4cm]{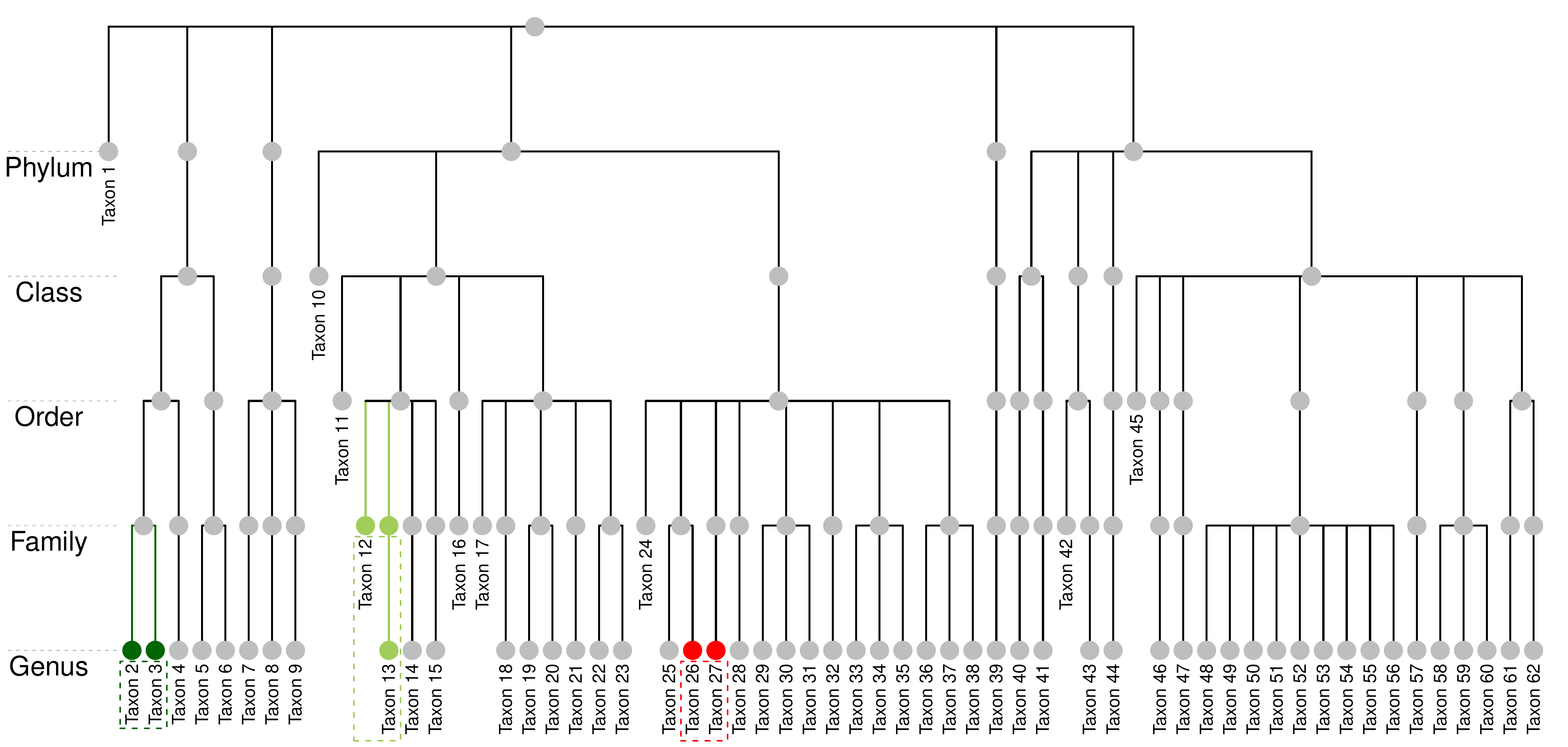}
       \caption{The NICU data: Taxonomic structure of the 62 taxa at the genus level. Taxon 2 and Taxon 3 share the same parent and are at the same taxonomic rank. Taxon 12 and Taxon 13 are at different ranks but they share the same lowest multi-child ancestor. Taxon 26 and Taxon 27 do not share the same lowest multi-child ancestor. 
       }\label{fig:weakstrong}
\end{figure*}

The raw microbiome data is longitudinal and has more than a thousand operational taxonomic units (OTU); these OTUs were classified up to the genus level using the RDP Classifier \citep{Cole2014}. For the purpose of illustration, we consider the average compositions over the postnatal period at the genus level, which results in a single dataset with $n=34$ subjects and $p=62$ taxa. Figure \ref{fig:taxonomy_nicu}(a) displays a heatmap of the data and \ref{fig:taxonomy_nicu}(b) shows the relative abundance barplot of the data. Figure \ref{fig:weakstrong} display the taxonomic tree of the 62 taxa up to the genus level.

To set up, suppose we observe $n$ independent compositional samples on $p$ taxa; let $\X = (\x_1, \ldots, \x_n)^T = (\tx_1, \ldots, \tx_p) = [x_{ij}]_{n\times p}$ be the observed $n\times p$ compositional data matrix, where each row $\x_i$, $i=1,\ldots,n$, lies in $\mathbb{S}^{p-1}$, with $\simp^{p-1}=\{\bx\in\real^p: \sum_{j=1}^p x_j=1, x_j\geq 0, j=1,\ldots,p\}$ representing a $(p-1)$-simplex in $\real^p$. As seen from the heatmap in Figure \ref{fig:taxonomy_nicu}, microbiome data is often very sparse with the presence of many rare taxa; even after being aggregated to the genus level, the percentage of zero entries in the data is still close to 40\%. 

In addition, we assume the availability of a taxonomic tree structure of the $p$ taxa. 
Some general terminologies of a tree structure are defined as follows. Let $T$ represent a $p$-leafed taxonomic
tree, $I(T)$ the set of internal nodes, and $|T|$ the total number of nodes in a
tree. Each leaf node of the tree corresponds to a taxon, and each internal node corresponds to a group of taxa. We follow the commonly used notions of child, parent, sibling, descendant, and ancestor to describe relations between nodes. Let the depth of a node $E$, denoted as $\mathcal{D}(E)$, be the number of ancestors from the node to the root, and let the depth of a tree, denoted as $\mathcal{D}^*(T)$, be the maximum depth of its leaf nodes. For a leaf node $E$, we use $A^*(E)$ to denote its lowest multi-child ancestor that has more than one child. For example, in Figure \ref{fig:weakstrong}, the depth of Taxon 1 is 1, while that of Taxon 2 is 5. The lowest multi-child ancestor of Taxon 12 is its parent, while that of Taxon 13 is its grandparent as its parent has only one child; that is, they share the same lowest multi-child ancestor. Taxon 26 and Taxon 27, on the other hand, do not share the same parent nor the lowest multi-child ancestor.

\section{Taxonomy-Guided Principal Amalgamation Analysis}\label{sec:PAA}

Analogous to PCA, which finds a number of principal components to best preserve the total variation in the data, PAA aims to aggregate the compositions to a smaller number of \textit{principal compositions}, guided by the available taxonomic structure, to best preserve a proper measure of information in the data.

In this section, we start with a general framework of PAA as an information-preserving optimization procedure in Section \ref{sec:PAA:framework}. To achieve scalable computation and conveniently utilize the taxonomic structure, a hierarchical agglomerative PAA (HPAA) algorithm is developed in \ref{sec:PAA:HPAA}. The computational details with various loss functions of interest are provided in Section \ref{sec:PAA:loss}, and the graphical tools for visualizations of PAA are illustrated in \ref{sec:PAA:graph}.

\subsection{Framework}\label{sec:PAA:framework}

The \textit{amalgamation} is a fundamental operation for compositions; it is formally defined as follows. 

\begin{definition}[Amalgamation]
  Let $\x \in \mathbb{S}^{p-1}$. For any $1 < k<p$, define
  \begin{align*}
    &\mathcal{M}_0(k,p) = \\
    &\left\{ \R = [r_{ij}]_{k\times p}; r_{ij} \in \{0,1\}, 
    \sum\nolimits_{i=1}^{k}r_{ij} = 1 \mbox{ for } j=1,\ldots,p\right\}.
  \end{align*}
Then, $\y = \R\x$ gives an amalgamation of $\x$ when $\R \in \mathcal{M}_0(k,p)$. 
\end{definition}

The matrix $\R = (\r_1,\ldots,\r_k)^T \in \mathbb{R}^{k\times p}$ is called an amalgamation matrix. It is clear that the $k$ many $\r_j$ vectors represent $k$ mutually exclusive and exhaustive subsets of the $p$ components in $\x$, and each $\r_j^T\x$ then computes a sum of the components of $\x$ in the $j$th subset. The operation of amalgamation reduces the original compositional vector in $\mathbb{S}^{p-1}$ to a simplex with dimension at most $k$, as it is possible that $\sum_{j=1}^{p}r_{ij}=0$ for some $i = 1,\ldots, k$. Apparently, this operation may result in a loss of information whenever $k < p$.

Naturally, given the observed data $\X$, PAA can be formulated as a set of optimization problems: for $k = 2,\ldots, p-1$, 
\begin{align}
  \widehat{\R}_{k} \in \arg\min_{\R \in \mathcal{M}_0(k,p)} L(\R;\X),
  \label{eq:paa1}
\end{align}
where $L(\cdot)$ is a properly specified loss function that measures certain information reduction from $\X$ to $\X\R^T$. Borrowing the terminology from PCA, we call the resulting $\widehat{\R}_{k}$ matrix as the {\it loading matrix} or the \textit{principal amalgamation matrix} and the amalgamated data $\X\widehat{\R}_{k}^T$ as the {\it score matrix}. Subsequently, the $k$ amalgamated vectors, $\X\widehat{\r}_1, \ldots, \X\widehat{\r}_k$ are called \textit{principal compositions}.

The construction of the loss function is flexible, and it is tied to the choice of how to measure the information in the data. For microbiome data or compositional data in general, of particular interest in practice is to measure the information loss by the reduction in some diversity index, for preserving a specific aspect of diversity in the original data as much as possible. Popular choices include the family of $\alpha$ diversity such as Simpson's diversity index (SDI) and Shannon–Wiener index (SWI), which measures within-sample diversity, and the family of $\beta$ diversity such as Bray-Curtis dissimilarity (BC) and Weighted UniFrac (WUF), which measures between-sample diversity \citep{Whittaker1960,Whittaker1972,Goodrich2014,Wagner2018}. There are also entropy-based or model-based measures that incorporate different aspects of the above indices \citep{renyi1961,Hill1973,Jost2006,Gotelli2013,RAJARAM201635}. Other methods for constructing the loss function include the likelihood approach, i.e., through making distributional assumption of the data (e.g., Dirichlet), and the transformation approach, i.e., through transforming the data to the Euclidean space such that familiar statistics such as sample variance can be used. To focus on the main idea, we defer the detailed discussions on these choices in Section \ref{sec:PAA:loss}.


However, due to the complex structure of $\mathcal{M}_0(k,p)$, conducting PAA through exactly solving the optimization problems in \eqref{eq:paa1} would be computationally prohibitive when $p$ is large, not even to mention that utilizing the taxonomic structure may introduce further complications. Intriguingly, we realize that PAA can be pursued from a cluster analysis perspective. With any specified number of principal compositions, the objective of PAA is essentially to search for a cluster pattern of the compositions, such that each cluster of compositions aggregates into a new principal composition.

\subsection{Hierarchical PAA Algorithm}\label{sec:PAA:HPAA}

Motivated by the connection between PAA and clustering, we develop an agglomerative hierarchical PAA (HPAA) algorithm to utilize taxonomic structure and enable scalable computation. Our approach starts from the original compositions and gradually amalgamates them through a sequence of simple amalgamations, i.e., at each step only a single pair of compositions is being aggregated. As such, HPAA generates the entire path of the simple amalgamations for reducing the data from its most informative original form to an utterly non-informative vector of ones. Along this process, a dendrogram of the successive amalgamations and the associated information losses is naturally generated.

We first describe the HPAA algorithm and the growth of the associated dendrogram in its basic form, without consideration of taxonomic guidance. To initialize, let $t=0$ and denote $\X_0 = \X = (\x_1, \ldots, \x_n)^T = (\tx_1, \ldots, \tx_p)$ as the original $n\times p$ compositional data matrix. We start from the $p$ taxa in the original data $\X_0$, each of which forms its own cluster. Let $\mathcal{S}_0 = \{\{1\},\ldots,\{p\}\}$ be the initial partition of the $p$ taxa, which correspondingly forms the initial leaf nodes of the dendrogram, denoted as $E_{0,1},\ldots, E_{0,p}$.

At the $t$th step, for $t=1,\ldots, p-1$, let $\mathcal{S}_{t-1}$ denote the set of $|\mathcal{S}_{t-1}|$ (i.e., $p-t+1$) nodes and $\X_{t-1}$ denote the $n\times |\mathcal{S}_{t-1}|$ current amalgamated data from the last step.
With these inputs, the core problem is to search for a pair of the current nodes, $(E_{t-1,\widehat{j}},E_{t-1,\widehat{j'}})$, to be aggregated into a new node such that the information loss of the amalgamated data is minimized. That is,
\begin{equation}
(\widehat{j}, \widehat{j'}) = \arg\min_{(j,j')\in \mathcal{P}_{t-1}} L(\R(j,j');\X_{t-1}),\label{eq:hpaa}
\end{equation}
where $\R(j,j')$ is a simple amalgamation matrix in $\mathcal{M}_0(|\mathcal{S}_{t}|,|\mathcal{S}_{t-1}|)$ that aggregates the $j$th and $j'$th columns of $\X_{t-1}$, and  
$\mathcal{P}_{t-1}$ is the active set of ``legitimate'' pairs of nodes that can be amalgamated. For instance, if no restriction is imposed, we set $\mathcal{P}_{t-1}=\{ (j,j'); 1\leq j<j' \leq |\mathcal{S}_{t-1}|\}$, consisting of all possible pairs of the current leaf nodes.

With the solution from \eqref{eq:hpaa}, we then update
\[
\X_{t} \leftarrow \X_{t-1}\R(\widehat{j},\widehat{j'})^T
\]
and denote the reduced set of nodes as $E_{t,1},\ldots, E_{t,p-t}$. For example, if at the first step ($t=1$), $E_{0,1}$ and $E_{0,2}$ are chosen to be combined, we have   $\mathcal{S}_1 = \{\{1,2\},\{3\},\ldots,\{p\}\}$, $\X_1 = (\tx_1 + \tx_2, \tx_3, \ldots, \tx_p)$, and the new reduced set of nodes are denoted as $E_{1,1},\ldots, E_{1,p-1}$.

The above procedure is repeated until only two nodes are left; they are then bound to be combined as a vector of ones. The proposed algorithm is summarized in Algorithm \ref{alg:unsup}.


\begin{algorithm}[h]
\caption{Hierarchical principal amalgamation analysis (HPAA) via agglomerative clustering}\label{alg:unsup}
\begin{algorithmic}[1]
\State \algorithmicrequire \ Compositional data $\X \in \mathbb{R}^{n\times p}$, and a user-specified loss function $L(\R;\X)$. 
\State \textbf{Initialization}: Set $\X_0 = \X$. Set the initial partition as $\mathcal{S}_0 = \{E_{0,j} = \{j\}, j = 1, \ldots, p\}$, where $E_{0,j} = \{j\}$ means the node/cluster $E_{0,j}$ is formed by the $j$th taxon only. 
\State \textbf{For} $t = 1, 2,\ldots, p-1$,
\begin{itemize}
\item Search for a pair of current nodes $E_{t-1,\widehat{j}}$ and $E_{t-1,\widehat{j'}}$ to perform amalgamation, by solving \eqref{eq:hpaa}.
\item Combine $E_{t-1,\widehat{j}}$ and $E_{t-1,\widehat{j'}}$ to be a new node, and accordingly update $\mathcal{S}_t$, $\X_t$ and $E_{t,j}$ ($j = 1,\ldots, p-t$).
\end{itemize}
\textbf{End For.}
\end{algorithmic}
\end{algorithm}



We propose three levels of taxonomy guidance: unconstrained, weak taxonomic hierarchy, and strong taxonomic hierarchy, which, as the names suggest, produce amalgamation patterns with different degrees of conformity with the taxonomic tree. It all boils down to properly set the active set of the paired nodes $\mathcal{P}_{t-1}$ in solving \eqref{eq:hpaa}. Moreover, when either weak or strong taxonomic hierarchy is enforced, the successive growth of the dendrogram through guided amalgamations is always coupled with the successive reduction of the taxonomic tree. 

\begin{itemize}

  \item \textit{Unconstrained}. In each step, we search over all possible pairs of nodes in solving \eqref{eq:hpaa},
\[
\mathcal{P}_{t-1}=\{ (j,j'); 1\leq j<j' \leq |\mathcal{S}_{t-1}|\}.
\]  
\item \textit{Weak taxonomic hierarchy}. In each step, we only search over pairs of nodes that share the same lowest multi-child ancestor in the reduced taxonomic tree. 
That is, \eqref{eq:hpaa} is solved over
  \begin{align*}
    &\mathcal{P}_{t-1} = \\
    &\{ (j,j'); 1\leq j<j' \leq |\mathcal{S}_{t-1}|, A^*(E_{t-1,j}) = A^*(E_{t-1,j'})\}.
  \end{align*}
For example, consider the first step of HPAA ($t=1$) with the $p=62$ leaf nodes in Figure \ref{fig:weakstrong}. Both (Taxon 2, Taxon 3) and (Taxon 12, Taxon 13) are in $\mathcal{P}_{0}$, while (Taxon 26, Taxon 27) is not.

\item \textit{Strong taxonomic hierarchy}. In each step, we further restrict the search to be among pairs of nodes that have the largest depth in the taxonomic tree. As a result, taxa at the lowest taxonomic rank will always be aggregated first. 
That is, \eqref{eq:hpaa} is solved over
\begin{align*}
  \mathcal{P}_{t-1} =\{ (j,j'); 
  & 1\leq j<j' \leq |\mathcal{S}_{t-1}|, \\
  & A^*(E_{t-1,j}) = A^*(E_{t-1,j'}), \\
  & \mathcal{D}(E_{t-1,j}) = \mathcal{D}(E_{t-1,j'}) = \mathcal{D}^*_{t-1}\}.
\end{align*}
For example, in Figure \ref{fig:weakstrong}, the pair (Taxon 2, Taxon 3) remains in $\mathcal{P}_{0}$, while (Taxon 12, Taxon 13) is no longer in $\mathcal{P}_{0}$ as they are not of the lowest taxonomic rank. 
\end{itemize}

\subsection{Construction of Loss Function with Common Diversity Measures}\label{sec:PAA:loss}

We illustrate the implementation of HPAA using loss functions constructed from several commonly-used $\alpha$ diversity and $\beta$ diversity measures.

The $\alpha$ diversity measures the richness (number of different entities) and evenness (the homogeneity in abundance of the entities) within each compositional sample. It can be calculated for each sample in the data, i.e., $\alpha(\x_i)$ for $i = 1, \ldots, n$. In general, larger $\alpha(\x_i)$ indicates larger within-sample diversity among species, and the index is non-increasing along successive amalgamations. Therefore, a general loss function based on $\alpha$ diversity can be constructed as
\begin{align}
  L_{\alpha}(\R;\X) = - \sum_{i=1}^{n}\alpha(\R\x_i).\label{eq:alpha}
\end{align}

\subsubsection*{Simpson's diversity index (SDI)}
Consider first the Simpson's diversity index (SDI), defined as
      \[
        \textrm{SDI}(\x_i) = 1- \sum\nolimits_{j=1}^{p}x_{ij}^2 = 1 - \x_i^T\x_i,
      \]
      where $x_{ij}$ represents the abundance of the $j$th components in the $i$th sample with $\x_i \in \mathbb{S}^{p-1}$. The SDI can be understood as the probability that two individuals randomly selected from a sample will belong to different species. A small SDI indicates that a few components dominate, while a large SDI indicates a diverse and balanced distribution among components. It is seen that SDI is non-increasing along successive amalgamations, as $x_{ij}^2 + x_{ij'}^2 \leq (x_{ij} + x_{ij'})^2$ for $x_{ij},x_{ij'}\geq 0$. Therefore, 
with the form of the loss function in \eqref{eq:alpha}, the general PAA criterion in \eqref{eq:paa1} becomes
      \[\min_{\R \in \mathcal{M}_0(k,p)} \mathrm{tr}(\R\X^T\X\R^T),\]
and the $t$th step simple amalgamation problem in \eqref{eq:hpaa} becomes
\[
\min_{(j,j')\in \mathcal{P}_{t-1}} \tx^T_{j,t-1}\tx_{j',t-1}, 
\]
where $\tx_{j,t=1}$ denotes the $j$th column of $\X_{t-1}$, which is equivalent to find the minimal off-diagonal element of $\X^T_{t-1}\X_{t-1}$ within the active set specified by $\mathcal{P}_{t-1}$. We remark that in each step only two columns of the amalgamated data are effected; this is utilized to simplify the computation.

\subsubsection*{Shannon–Wiener index (SWI)}
Unlike the SDI which weights more on dominant components, the Shannon–Wiener index (SWI) is equally sensitive to rare and dominant components, defined as
      \[
        \textrm{SWI}(\x_i) = - \sum\nolimits_{j=1}^{p} x_{ij} \log x_{ij} = - \x_i^T \log(\x_i).
      \]
The logarithmic transformation is applied entrywisely on the enclosed vector or matrix. As such, to compute SWI, we do need to first replace zeros in the data. The SWI is non-increasing along successive amalgamations since $x_{ij} \log x_{ij} + x_{ij'} \log x_{ij'}\le (x_{ij} + x_{ij'}) \log (x_{ij} + x_{ij'})$ for $x_{ij}, x_{ij'} > 0$. Therefore, 
with the loss function form in \eqref{eq:alpha}, the general PAA criterion in \eqref{eq:paa1} becomes
      \[
        \min_{\R \in \mathcal{M}_0(k,p)} \mathrm{tr}\{\R\X^T\log(\X\R^T)\},
      \]
and the $t$th step simple amalgamation problem in \eqref{eq:hpaa} becomes
\begin{align*}
\min_{(j,j')\in \mathcal{P}_{t-1}} 
\{&(\tx_{j,t-1} + \tx_{j',t-1})^T\log(\tx_{j,t-1} + \tx_{j',t-1}) \\
  & - \tx_{j,t-1}^T\log(\tx_{j,t-1}) - \tx_{j',t-1}^T\log(\tx_{j',t-1})\}.
\end{align*}

While the $\alpha$ diversity focuses on within-sample diversity, the $\beta$ diversity reflects the between-sample differences. It can be calculated for each pair of samples in the data, i.e., $\beta(\x_i, \x_{i'})$ for $i, i' = 1, \ldots, n$, resulting in a between-sample distance or dissimilarity matrix $\bD(\X) = [\beta(\x_i, \x_{i'})]_{n\times n}$.
As such, PAA aims to best preserve the dissimilarity pattern in the amalgamated data. 
We thus construct the loss function based on $\beta$ diversity as the sum of squared differences between the original distance matrix and that of the amalgamated data,
    \begin{align}\label{eq:beta-diversity}
      L_\beta(\R;\X)
      = \sum\nolimits_{i < i'} \{\beta(\x_i, \x_{i'}) - \beta(\R\x_i, \R\x_{i'})\}^2.
    \end{align}

\subsubsection*{Bray-Curtis dissimilarity index (BC)}
Consider the Bray-Curtis dissimilarity index (BC) defined as
      \begin{align*}
        \textrm{BC}(\x_i, \x_{i'})
        & = \frac{\sum\nolimits_{j=1}^p|x_{ij} -
        x_{i'j}|}{\sum\nolimits_{j=1}^p (x_{ij} + x_{i'j})}
        = \frac{\sum\nolimits_{j=1}^p |x_{ij} - x_{i'j}|}{2} \\
        &= \frac{\sum\nolimits_{j=1}^p [(x_{ij} + x_{i'j}) - 2\min(x_{ij}, x_{i'j})]}{2} \\
        &= 1 - \sum\nolimits_{j=1}^p \min(x_{ij}, x_{i'j}), 
      \end{align*}
      for any pair of samples $\x_i, \x_{i'} \in \mathbb{S}^{p-1}$. It is non-increasing along successive amalgamations as $\min(x_{ij}, x_{i'j}) + \min(x_{ij'}, x_{i'j'}) \le \min(x_{ij} + x_{ij'}, x_{i'j} + x_{i'j'})$.
The $t$th step simple amalgamation problem in \eqref{eq:hpaa} can be expressed as
\begin{align*}
  \min_{(j,j')\in \mathcal{P}_{t-1}}\{\sum\nolimits_{i < i'}[
  & \min(x_{ij,t-1},x_{i'j,t-1}) \\
  & + \min( x_{ij',t-1}, x_{i'j',t-1}) \\
  & - 
  \begin{aligned}[t]
  \min( &x_{ij,t-1} + x_{ij',t-1}, \\
        &x_{i'j,t-1} + x_{i'j',t-1})]^2\}.
  \end{aligned}
\end{align*}

\subsubsection*{Weighted UniFrac distance (WUF)}
The weighted UniFrac distance (WUF) further incorporates information from the phylogenetic or taxonomic tree when computing the between-sample distance. It can also be viewed as a plug-in estimate of the Wasserstain distance between two probability distributions defined on the taxonomic tree \citep{Evans2012phylo}. 
It is commonly used in exploratory microbiome data analysis and a number of variants were developed. To mention a few, double principal coordinate analysis (DPCoA) proposed by  Pavoine et al. \cite{Pavoine2004diss} generalized PCA by incorporating the relationship among variables from the phylogenetic structure that can be described using dissimilarity measures like UniFrac or weighted UniFrac. Chen et al. \cite{Chen2012unifrac} compared the power of statistical tests using a number of variants of UniFrac including unweighted/weighted UniFrac and
generalized UniFrac. Randolph et al. \cite{randolph2018kernel} proposed a kernel-based regression framework that incorporates the unweighted/weighted UniFrac dissimilarity matrix from the phylogenetic structure.

Here we briefly illustrate HPAA with weighted UniFrac distance. For any pair of samples $\x_i, \x_{i'} \in \mathbb{S}^{p-1}$, WUF is defined as
      \[
        \textrm{WU}(\x_i, \x_{i'}) = \frac{\sum\nolimits_{j=1}^p l_j |x_{ij} - x_{i'j}|}{\sum\nolimits_{j=1}^p L_j(x_{ij} + x_{i'j})},
      \]
where $l_j$ for $j=1,\ldots,p$ denotes the length of the $j$ branch, i.e., the length between the node for $j$th entity and its parent, and $L_j$ denotes the distance of $j$th entity from the root node of the phylogenetic tree. Here the length of branches may change with compositions at lower levels of taxonomic tree amalgamated to higher level.

At the $t$th step, the simple amalgamation problem in \eqref{eq:hpaa} can be expressed as
      \begin{align*}
        &\min_{(j,j')\in \mathcal{P}_{t-1}}\{\sum\nolimits_{i < i'} [
        \frac{\sum\nolimits_{j=1}^p l_j |x_{ij} - x_{i'j}|}{\sum\nolimits_{j=1}^p L_j(x_{ij} + x_{i'j})} -\\
        &\frac{\sum\nolimits_{k=1, k \ne j, j'}^p l_k |x_{ik} - x_{i'k}| + l_{j,j'}|(x_{ij} + x_{ij'}) - (x_{i'j} + x_{i'j'})|}{\sum\nolimits_{k=1, k \ne j. j'}^p L_k(x_{ik} + x_{i'k}) + L_{j,j'}(x_{ij} + x_{ij'} + x_{i'j} + x_{i'j'})}]^2\},
      \end{align*}
      where $l_{j,j'}$ denotes the the length between the newly formed entity from $j$th and $j'$th entities and its parent, and $L_{j,j'}$ denotes the distance of new entity from the root node of the phylogenetic tree. During the successive amalgamations, the lengths of branches in computing WUF are also getting updated. 

\subsection{Visualization with Examples}\label{sec:PAA:graph}

We use the NICU data to illustrate the graphical tools developed for visualizing the PAA results. These tools can be extremely useful for visualizing and understanding compositional data, as well as helping to determine the desired number of principal compositions in practice.

\subsubsection*{Dendrogram}
\begin{figure*}[!h]
\centering
       \includegraphics[width=11.4cm]{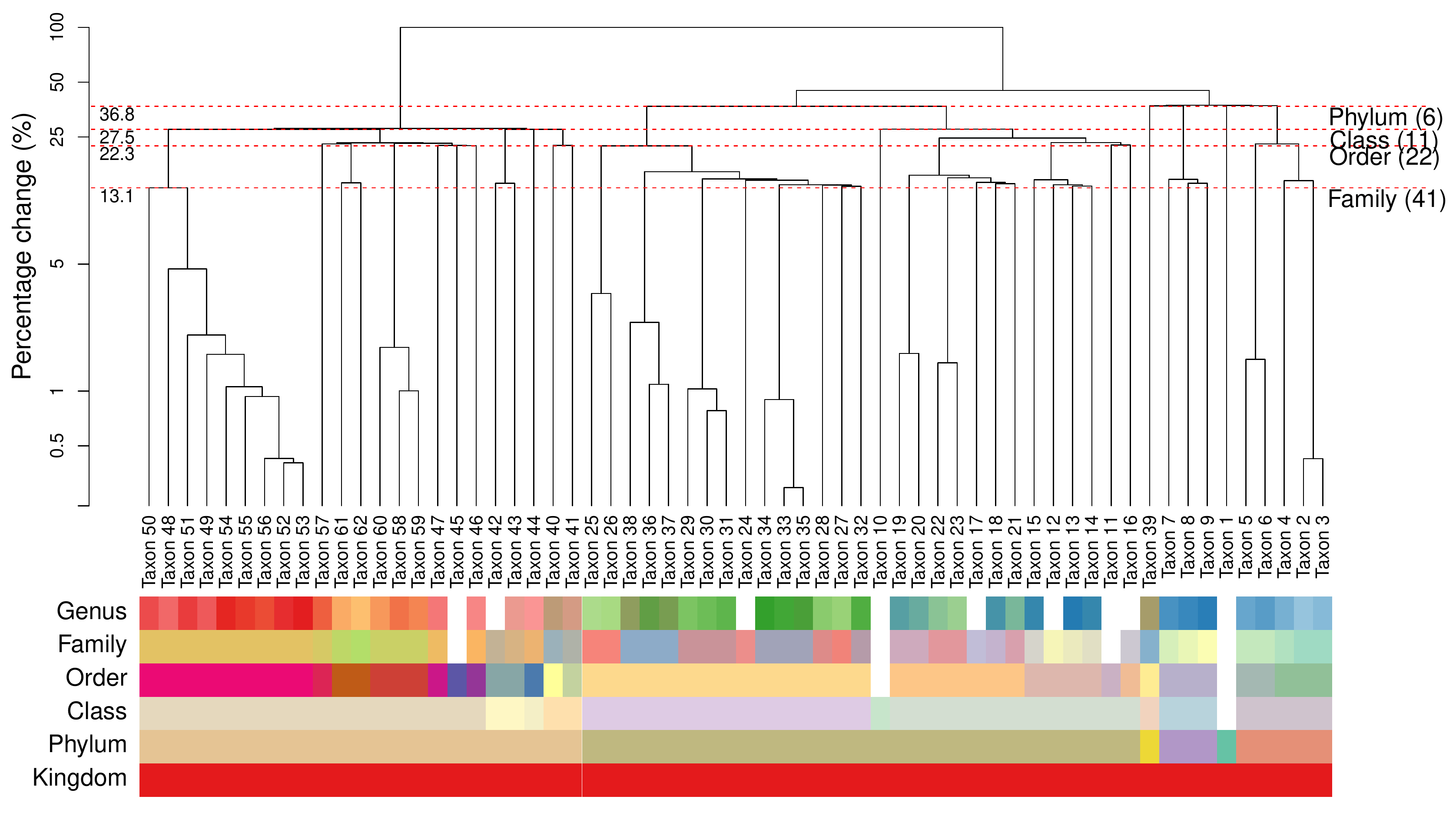}
\caption{The NICU data: Dendrogram of HPAA with SDI and strong taxonomic hierarchy.}\label{fig:dendrogram}
\end{figure*}

We construct a HPAA dendrogram to simultaneously visualize both the tree diagram of the successive amalgamations and the taxonomic structure of the taxon. To illustrate, Figure \ref{fig:dendrogram} shows the HPAA dendrogram from performing HPAA with SDI loss and strong taxonomic hierarchy on the NICU data. The top part of figure shows the dendrogram of amalgamations, where the $y-$axis shows the percentage decrease in total diversity as measured by SDI (on the log-scale) along the successive amalgamations, from the bottom to the top. As such, any horizontal cut of the dendrogram at a desired level of diversity loss/preservation shows the corresponding amalgamated data. In particular, each red dashed horizontal line indicates the steps at which the original data are aggregated to a higher taxonomic rank. It shows that, for example, aggregating data to the order level (22 taxa or principal compositions left) through HPAA leads to 22.3\% loss in total SDI. At the bottom part, we use color bars to show taxonomic structure of the taxa (as shown in Figure \ref{fig:taxonomy_nicu}), where in each horizontal bar taxa of the same color belong to the same category of that rank.

Figures \ref{fig:dendro_simpson_case1} and \ref{fig:dendro_BC_case1}  in Supplementary Information \ref{sec:supp:visual} show HPAA dendrograms with SDI loss and BC loss, respectively, under all three levels of taxonomy guidance. Not surprisingly, the patterns of amalgamations vary under different settings. Without taxonomic constraint, the change in diversity appears to be very smooth along the amalgamations, but the resulting principal compositions may not be easily interpretable, as indicated by the mixed color patterns in the color bars of the taxonomic rank. On the other hand, for the setting of strong taxonomic hierarchy, while the principal compositions are forced to closely follow the taxonomic structure, the percentage change in diversity tends to exhibit dramatic jumps, especially at the steps that the last remaining taxon at a lower taxonomic rank is forced to be aggregated to a higher rank. As a compromise, for the setting of weak taxonomic hierarchy, the resulting principal compositions remain interpretable, and the percentage change in diversity remains smooth and can be quite close to that of the unconstrained setting in the early stage of amalgamations.



The HPAA dendrogram also reveal several interesting insights on the microbiome of preterm infants. As shown in Figure~\ref{fig:dendrogram}, while Taxa 49-56 are all genus of the Enterobacteriaceae family, the pattern of amalgamation suggests that Taxon 50, which is Klebsiella, is distinctive with the rest. It turns out that Klebsiella is a genus of Enterobacteriaceae that has emerged as a significant nosocomial pathogen in neonates \citep{Hervas2001,Gupta2002}, and its species have been implicated as a cause of various neonatal infections \citep{Sood1989,Basu2001} and neonatal sepsis \citep{Podschun1998,Westbrook2000}.

\subsubsection*{Scree plot} 
\begin{figure*}[h]
\centering
\includegraphics[width=\textwidth]{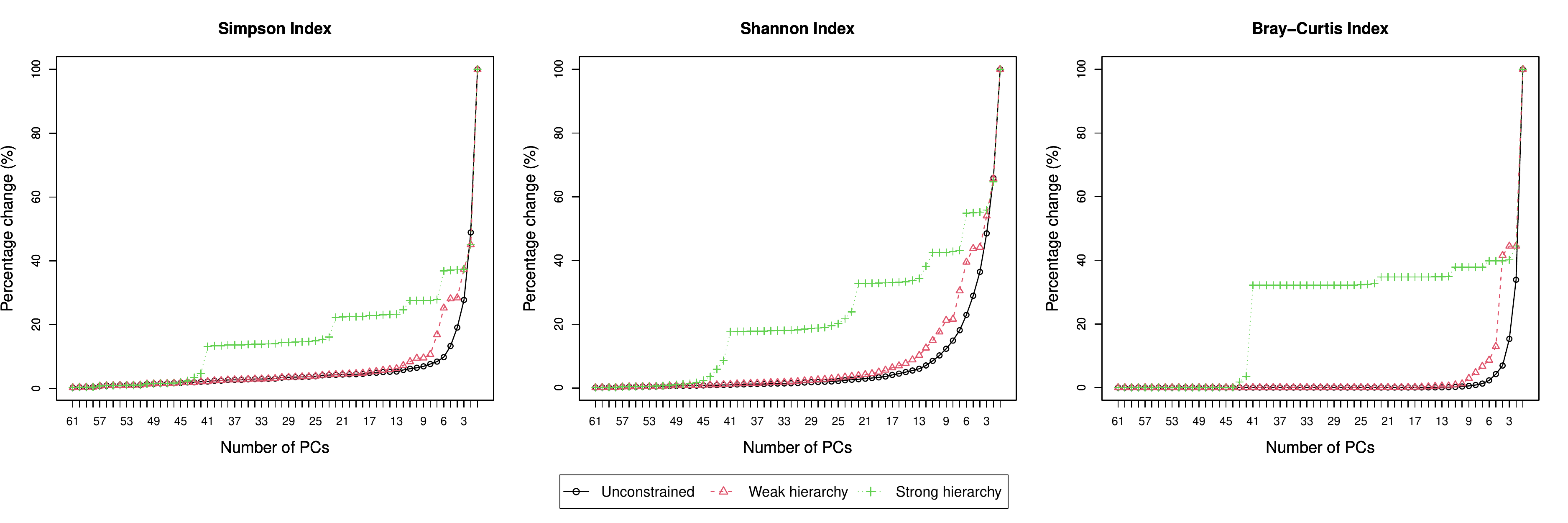}
\caption{The NICU data: Scree plots for HPAA (Percentage change in diversity vs. number of principal compositions).}\label{fig:percent_line_case1}
\end{figure*}

The scree plot shows the percentage change in the diversity loss as a function of the number of principal compositions. Figure \ref{fig:percent_line_case1} shows the scree plots from performing HPAA on the NICU data under different settings. The difference among the three levels of taxonomic guidance is very revealing, which confirms the previous observation from the dendrograms that the setting of weak taxonomic hierarchy reaches a good balance between preserving information and interpretability. 

\subsubsection*{Ordination plot} We construct ordination plot to visualize the changes in the between-sample distance patterns before and after HPAA with any given number of principal compositions. Specifically, we perform the non-metric multidimensional scaling (NMDS) analysis with Bray–Curtis dissimilarity on the combined original data and the principal compositions from HPAA, which produces a low-dimensional ordination plot of all samples before and after amalgamation. For each sample, it is represented by a pair of points from either the original data or the principal compositions; the smallest circle that covers the pair is drawn, whose radius then indicates the level of distortion due to HPAA data reduction.  

\begin{figure*}[!h]
\centering
\includegraphics[width=\textwidth]{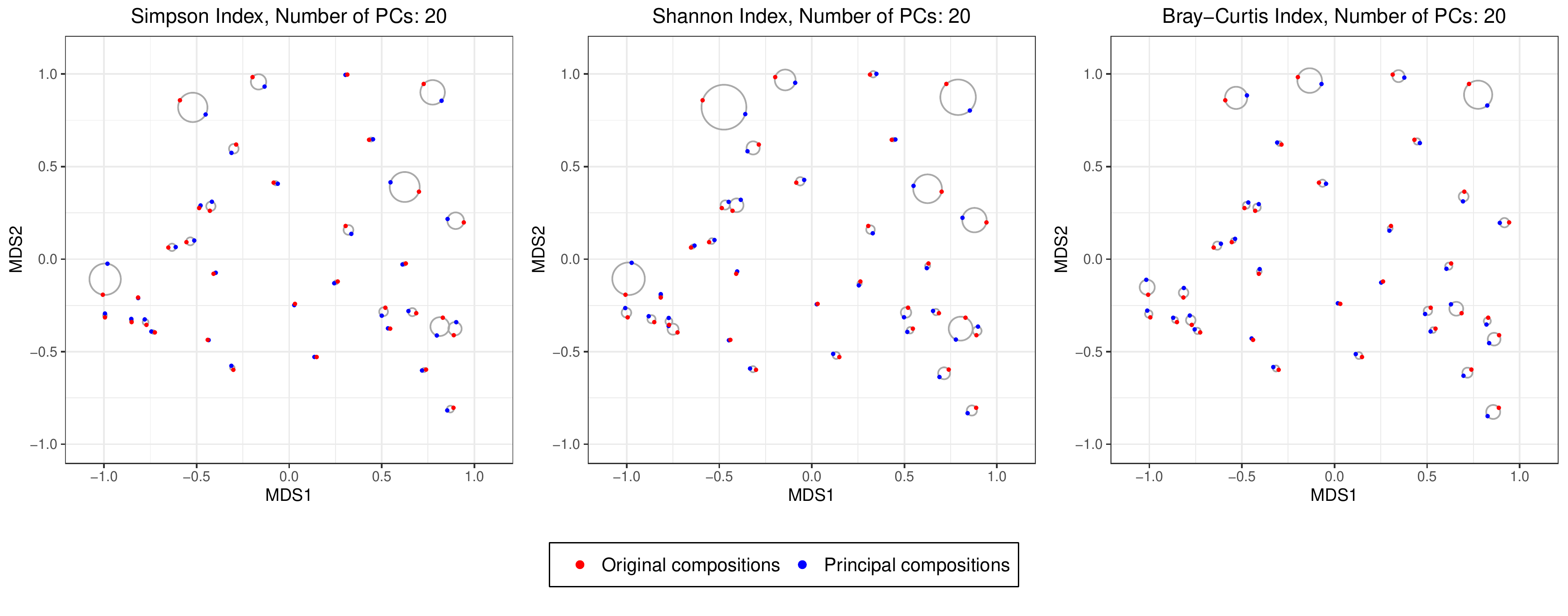}
\caption{The NICU data: 2D NMDS ordination plots for comparing original and principal compositions from HPAA with weak taxonomic hierarchy.}\label{fig:MDS_case1}
\end{figure*}

Figure~\ref{fig:MDS_case1} shows the ordination plots from performing HPAA on the NICU data with three different loss functions and weak taxonomic hierarchy, in which 20 principal compositions are kept. All three settings preserve the between-sample diversity reasonably well, as indicated by the fact that the circles generally have a small radius; as expected, HPAA with the BC loss performs the best as it directly targets on preserving between-sample diversity.

Figure~\ref{fig:MDS_BC_case1} in Supplementary Information \ref{sec:supp:visual} shows the ordination plots from performing HPAA on the NICU data with the BC loss and weak taxonomic hierarchy, with different numbers of principal compositions. As expected, the larger the number of principal compositions, the better the preservation of the between-sample diversity and the less reduction of the size of the data. In practice, such plots, together with the associated statistics, could be very useful in determining the appropriate number of principal compositions. 

\section{Comparison with Competing Methods}\label{sec:numeric}

To the best of our knowledge, the most relevant competitor of HPAA is the approach proposed by Quinn and Erb \cite{Quinn2020Amalgam}; it is implemented in an R package \texttt{amalgam} that is available from GitHub repository amalgam (\url{https://github.com/tpq/amalgam}). For any prespecified dimension of the amalgamated data, the method, referred to herein as \texttt{amalgam}, maximizes the correlation between the two Euclidean distance matrices computed from centered log-ratio transformed data before and after amalgamation. Due to the combinatory and nonconvex nature of the problem, a genetic algorithm was proposed to conduct local optimization. Besides \texttt{amalgam}, we also consider naive prevalence-based filtering method that simply discard taxa with low abundance.

\subsection{Simulation}

We first compare the computation efficiency of HPAA and \texttt{amalgam}. The results show that HPAA is very computationally efficient and scales well with the increase of the dimension or the sample size. In contrast, \texttt{amalgam} is very computationally intensive even for moderately large $p$ or $n$, making it unsuitable for large-scale microbiome studies. We also compare different dimension reduction methods on how well they preserve the between-sample distance pattern, which is very importance in many biological applications. The results show that HPAA methods with different loss functions outperform the baseline and the \texttt{amalgam} methods. PAA with the Bray-Curtis loss performs the best, as it directly aims at preserving the Bray-Curtis dissimilarity. To our surprise, the \texttt{amalgam} method performs even worse than the baseline, which may be due to its requirement of zero-replacement and log-ratio transformation and the slow convergence of their genetic algorithm. See details in Supplementary Information \ref{sec:supp:sim}.

\subsection{Application: Microbiome and HIV Infection}
\label{sec:app-case2}
  
Understanding the association between microbiome richness and HIV-1 risk may help to design novel
interventions to improve HIV-1-associated immune dysfunction. Here we considered a cross-sectional
HIV microbiome study conducted in Barcelona, Spain, that included both HIV-infected subjects and
HIV-negative controls \citep{Marc2016Gut}. Gut microbiome data were obtained from
MiSeq 16S rRNA sequence data on fecal microbiomes and bioinformatically processed with mothur. The
main goal of the study is to find the association between HIV transmission group (MSM: men who have sex with men vs non-MSM), HIV infection status and relative abundance of microbiome composition. As reported by Noguera-Julian et al. \cite{Marc2016Gut}, risk factors related with sexual preference such as MSM and non-MSM might greatly affect the gut microbiome composition, and thus the relative abundance of taxa might be able to identify the risk clusters of subjects. Following Quinn and Erb \cite{Quinn2020Amalgam}, the microbiome abundance data were preprocessed to produce a genus-level relative abundance data matrix for $p = 60$ taxa and $n=128$ HIV-infected subjects, including 60 MSM and 55 non-MSM subjects. The percentage of zeros is 36.6\%. The taxonomic tree structure of the $p=60$ taxa was also available as extrinsic information.

With this dataset, we compared different dimension reduction methods in terms of their performance on preserving the between-sample distance and on the classification accuracy of the MSM factor of subjects with the reduced data. For \texttt{amalgam}, the number of amalgamations is fixed at $k=20$. To be comparable, we use HPAA with BC loss and weak taxonomic hierarchy to produce $k=20$ principal compositions. We also include a simple prevalence-based filtering method that only keeps the $k=20$ taxa with the highest prevalence.

\begin{figure*}[t]
\centering
\includegraphics[width=\textwidth]{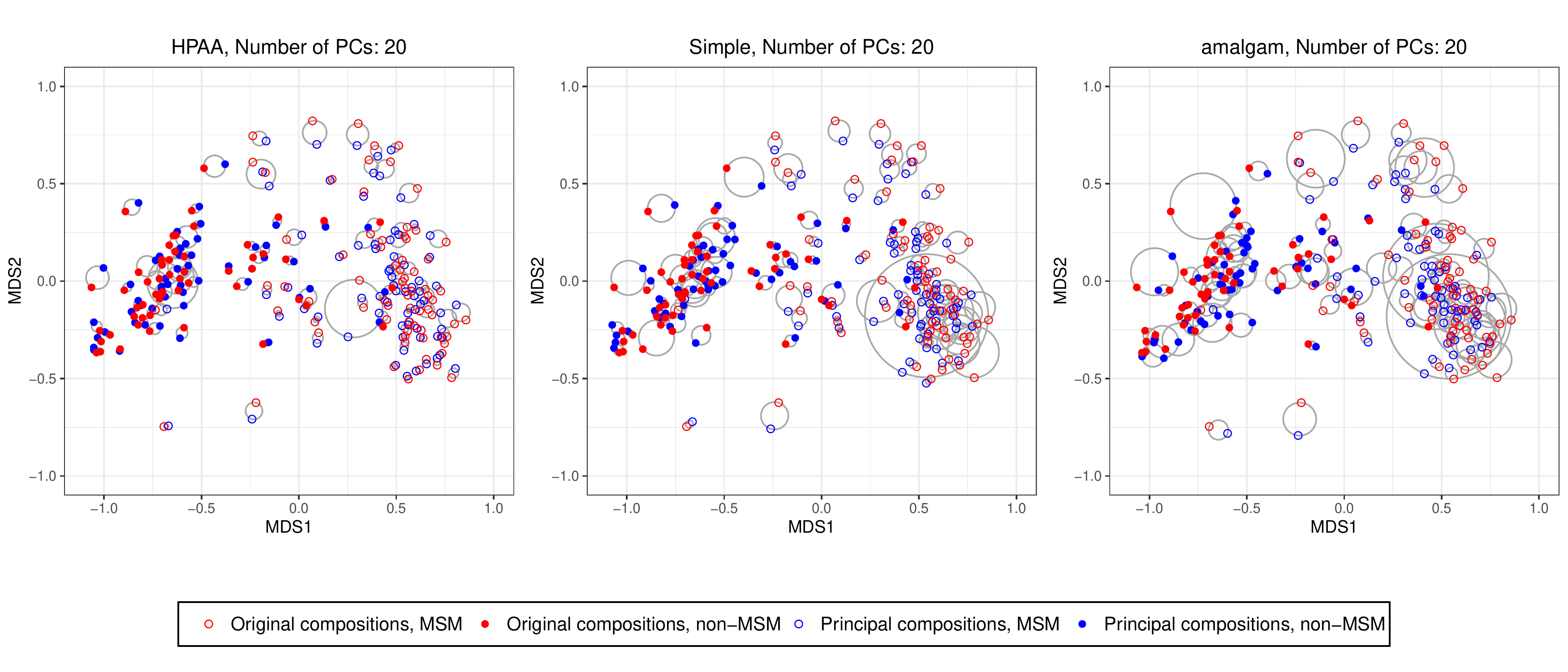}
\caption{The HIV data: NMDS ordination plots for comparing original data and different principal compositions from different dimension reduction methods including HPAA with Bray-Curtis and weak taxonomic hierarchy (HPAA), simple approach (Simple), and the \texttt{amalgam} method by Quinn and Erb \cite{Quinn2020Amalgam}.}\label{fig:MDS_comp_case2}
\end{figure*}

Figure~\ref{fig:MDS_comp_case2} shows the ordination plots of the three methods. The average Euclidean distance (with standard deviation in brackets) between the points representing the original compositions and principal compositions are 0.05 (0.04), 0.09 (0.07), 0.11 (0.09) for HPAA, the naive method, and the amalgam method, respectively. It is revealing that HPAA performs the best in preserving the between-sample distances of the data, which is partly owing to the proper use of the taxonomic structure.

To evaluate the performance of the three dimension reduction methods in downstream classification analysis, we conduct an out-of-sample random splitting procedure. In each run, we random split the original data into a training set of 80\% samples and a testing set of 20\% sample. Each of the three methods is applied on the training data to produce $k=20$ features, which are then used as predictors to train a logistic regression model of MSM status. The trained feature construction approach and the logistic regression model are then applied to the testing data, for which the classification performance is measured by the value of the area under the receiver operating characteristic curve (AUC). This procedure is 100 times. 
The average AUC values are 0.84 (0.07), 0.83 (0.08), and 0.82 (0.07) for HPAA, the naive method, and the amalgam method, respectively. Again, we see that the principal compositions from HPAA leads to improved classification, which showcases the potential of HPAA in improving downstream statistical analysis.


\section{Discussion}\label{sec:dis}

We have developed a new approach, principal amalgamation analysis, to perform dimension reduction of microbiome compositional data. The proposed method aggregates the compositions to a smaller number of principal compositions, by minimizing a user-specified loss function subject to conformity to the taxonomic structure. We hope to advocate using it as a preprocessing tool to reduce the dimension of highly-sparse OTU-level relative abundance data. 


\vspace{6pt}

\clearpage
\section*{Supplementary Materials}

\setcounter{figure}{0}
\setcounter{table}{0}
\renewcommand{\thefigure}{S\arabic{figure}}
\renewcommand{\thetable}{S\arabic{figure}}

\appendix
\section{Existing Methods}\label{sec:supp:literature}


Microbiome data are often normalized as compositions \citep{gloor2016s,tsilimigras2016compositional}, which reside in a simplex that does not admit the standard Euclidean geometry. It creates significant challenges for statistical analysis, as many standard methods do not directly apply.
There have been developments on compositional data analysis based on transformations and the so-called Aitchison geometry \citep{aitchison1982statistical,aitchison2005compositional,aitchison1983principal,aitchison1984log,Bacon2011}. 
However, these transformations could be inadequate to accommodate the unique features of microbiome data such as zero inflation, over dispersion and the presence of taxonomic tree structure among microbes.

The existing data reduction approaches for microbiome compositional data can be summarized to four categories, namely, the indexing approach, the selection approach, the transformation-based approach, and the amalgamation-based approach. The indexing approach, which represents data by some diversity or complexity indices, typically disables taxon-level analysis and results in oversimplification of data \citep{Johnson2016,Wagner2018,Willis2019}. The selection approach \citep{ChenLi2013,Susin2020}, which only keeps a subset of ``dominant'' compositions, often ignores intrinsic relations due to compositionality and can be vulnerable to the extremely low abundances of many OTUs. In practice, such selection could trivially end up with a few most prevalent ones \citep{Aitchison1984reducing}. The transformation-based approach conveniently utilizes existing reduction methods such as PCA after transforming data to the Euclidean space \citep{aitchison2005compositional,ZouHastie2006,filzmoser2009principal,scealy2015robust,wang2015principal}. The required transformations usually involve logarithm operations and cannot be directly applied on the excessive zeros in the microbiome data. An alternative is to use power transformations such as square-root transformation \citep{scealy2015robust,Dai2018}, which avoids zero replacement and puts the data onto the unit sphere to enable manifold-based PCA. However, these PCA approaches may compromise interpretation of the data in terms of individual taxon and impede incorporation of the extrinsic taxonomic tree structure. Other proximity-based methods such as PCoA \citep{AndersonMJ2003,Verma2020} could accommodate several special features of the data but fail to pinpoint specific taxon that drive data reduction. 

While Aitchison's formal terminology of ``amalgamation'' may not be as widely spread as it should be, the operation itself has nevertheless been widely used in microbiome data analysis, although often as a pragmatic and rather ad-hoc way of dealing with the most rare compositional components in the data. For example, rare taxon with excessive number of zeros or low abundances at lower taxonomic ranks are aggregated to a higher rank for analysis \citep{lin2014variable, randolph2018kernel}. It is also common in the microbiome analysis that rare taxon are simply removed by some ad-hoc filtering process \citep{Cao2021rare}. These naive approaches may lead to unwanted information loss and potential conflicts between analyses performed at different ranks or with different filter rules.

Not until recently, a few studies on amalgamation-based dimension reduction emerged. Greenacre \cite{Greenacre2020Amalgam} and  Greenacre et al. \cite{Greenacre2021Amalgam} argued that amalgamation provides an interpretable way to reduce the dimensionality of compositions and could make substantive sense in practical applications, despite the non-linearity in the Aitchison geometry of the simplex and its possibility to distort between-sample distances. They advocated for expert-driven amalgamation, i.e., the use of domain-knowledge to perform amalgamation, and proposed amalgamation-based clustering with log-ratio transformed data. Quinn and Erb \cite{Quinn2020Amalgam} further discussed the usage of amalgamation as an alternative to the commonly-adopted dimension reduction methods and proposed an optimization approach to preserve a suitable between-sample distance measure with centered log-ratio transformation. However, the method does not work without zero-replacement, and their genetic algorithm can be extremely computational intensive and hinders the incorporation of extrinsic information such as the taxonomic tree structure.

\clearpage
\section{Visualization Example}\label{sec:supp:visual}

\begin{figure*}[htp]
\centering
\includegraphics[width=0.85\textwidth]{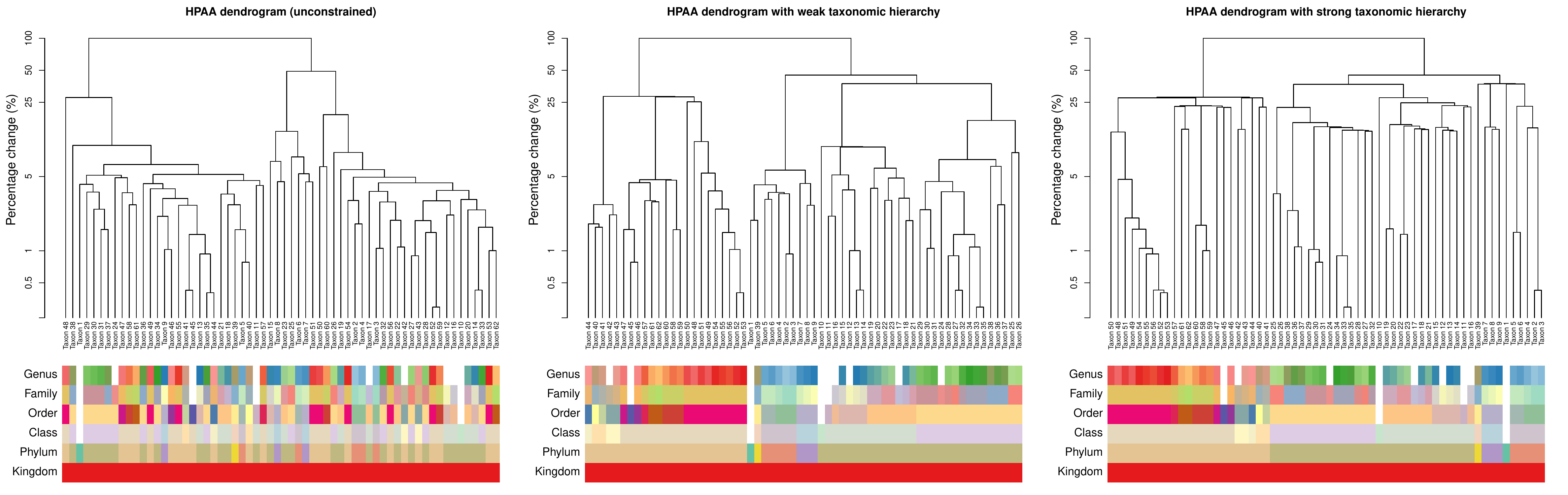}
\caption{The NICU data: HPAA dendrograms with SDI and different
  constraints on taxonomic hierarchy.}\label{fig:dendro_simpson_case1}
\end{figure*}

\begin{figure*}[h]
  \centering
  \includegraphics[width=0.85\textwidth]{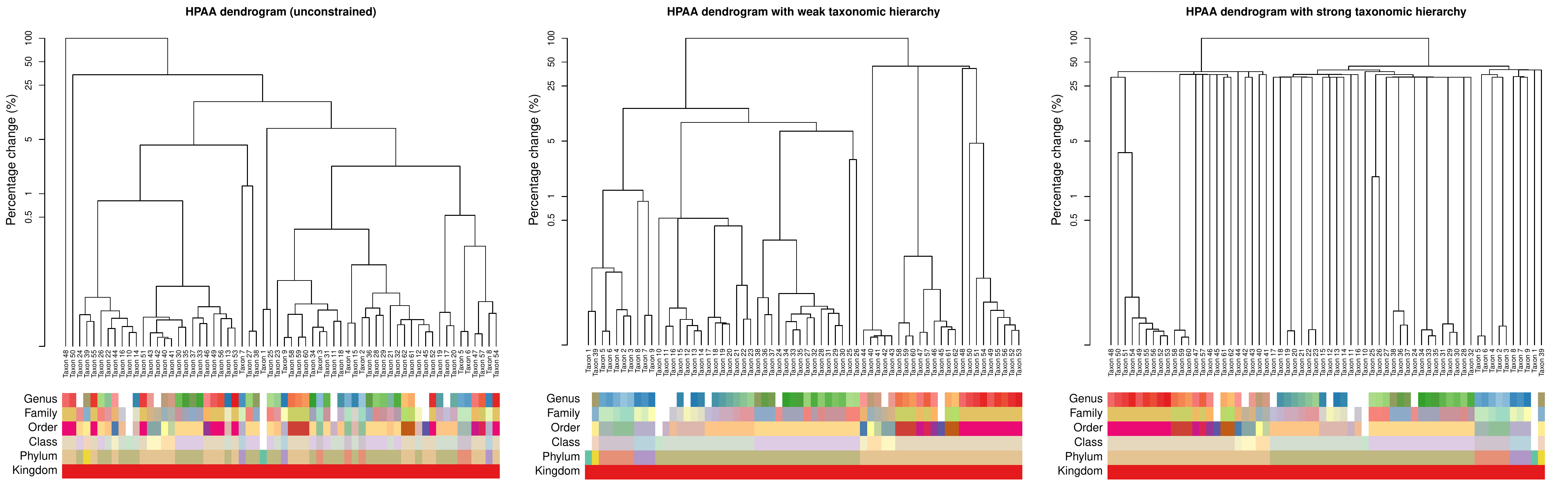}
  \caption{The NICU data: HPAA dendrograms with Bray-Curtis and
    different constraints on taxonomic
    hierarchy.}\label{fig:dendro_BC_case1}
\end{figure*}

\begin{figure*}[!h]
  \centering
  \includegraphics[width=0.85\textwidth]{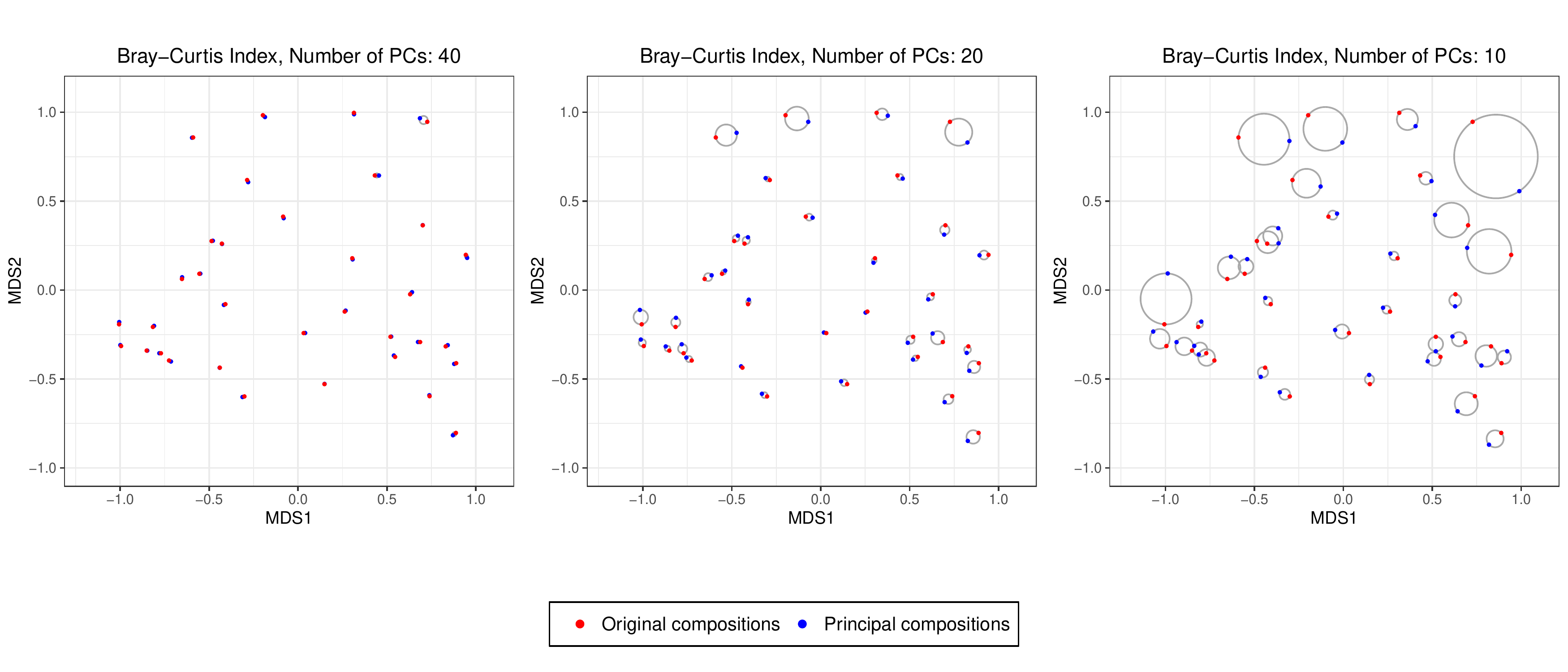}
  \caption{The NICU data: 2D NMDS ordination plots for comparing
    original data and different numbers of principal compositions from
    HPAA with Bray-Curtis and weak taxonomic
    hierarchy.}\label{fig:MDS_BC_case1}
\end{figure*}

\clearpage
\section{Simulation}\label{sec:supp:sim}

\begin{figure*}[htp]
  \centering
  \subfigure[Effect of dimension]{\includegraphics[width=0.44\textwidth]{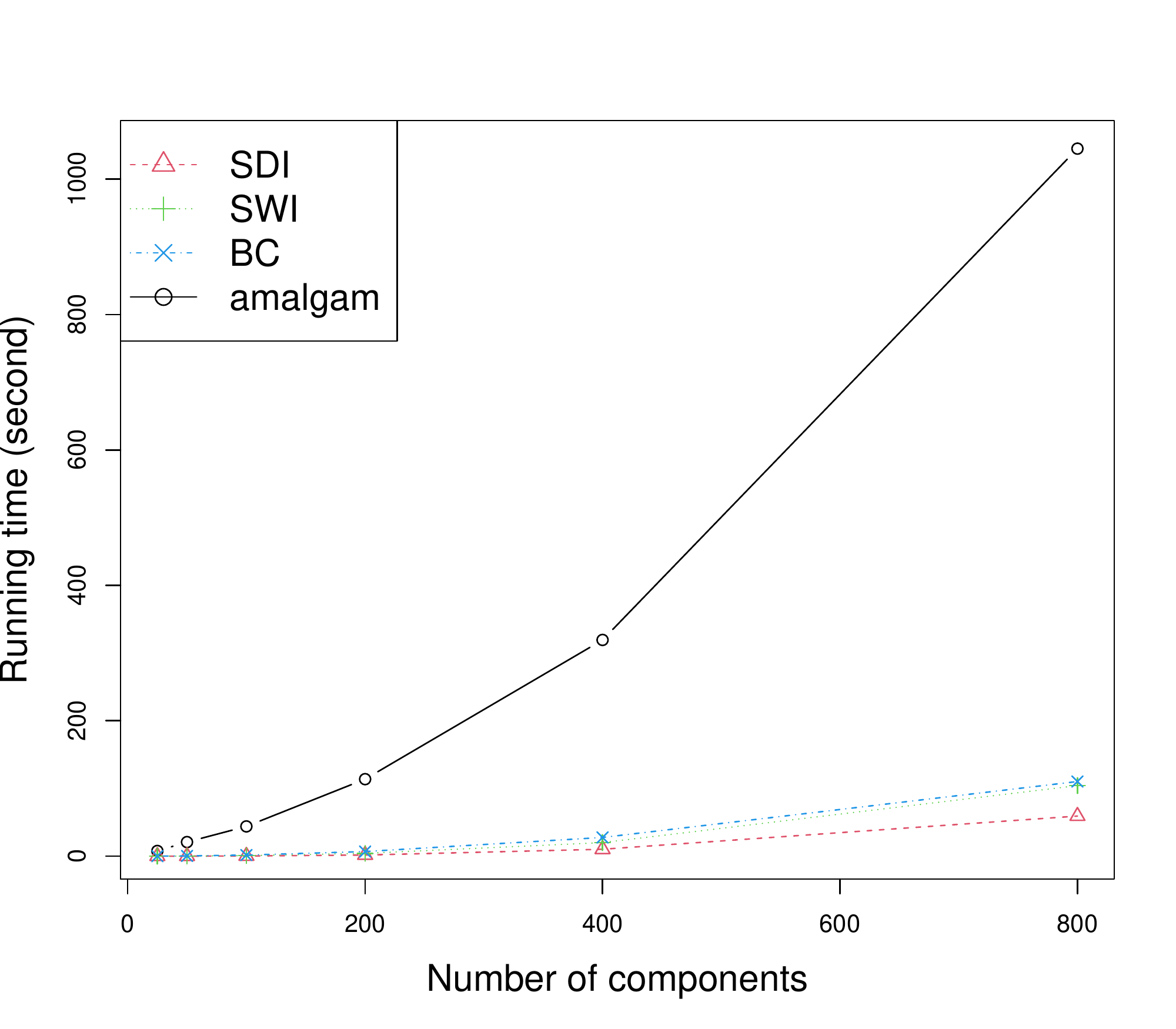}}
  \subfigure[Effect of sample size]{\includegraphics[width=0.44\textwidth]{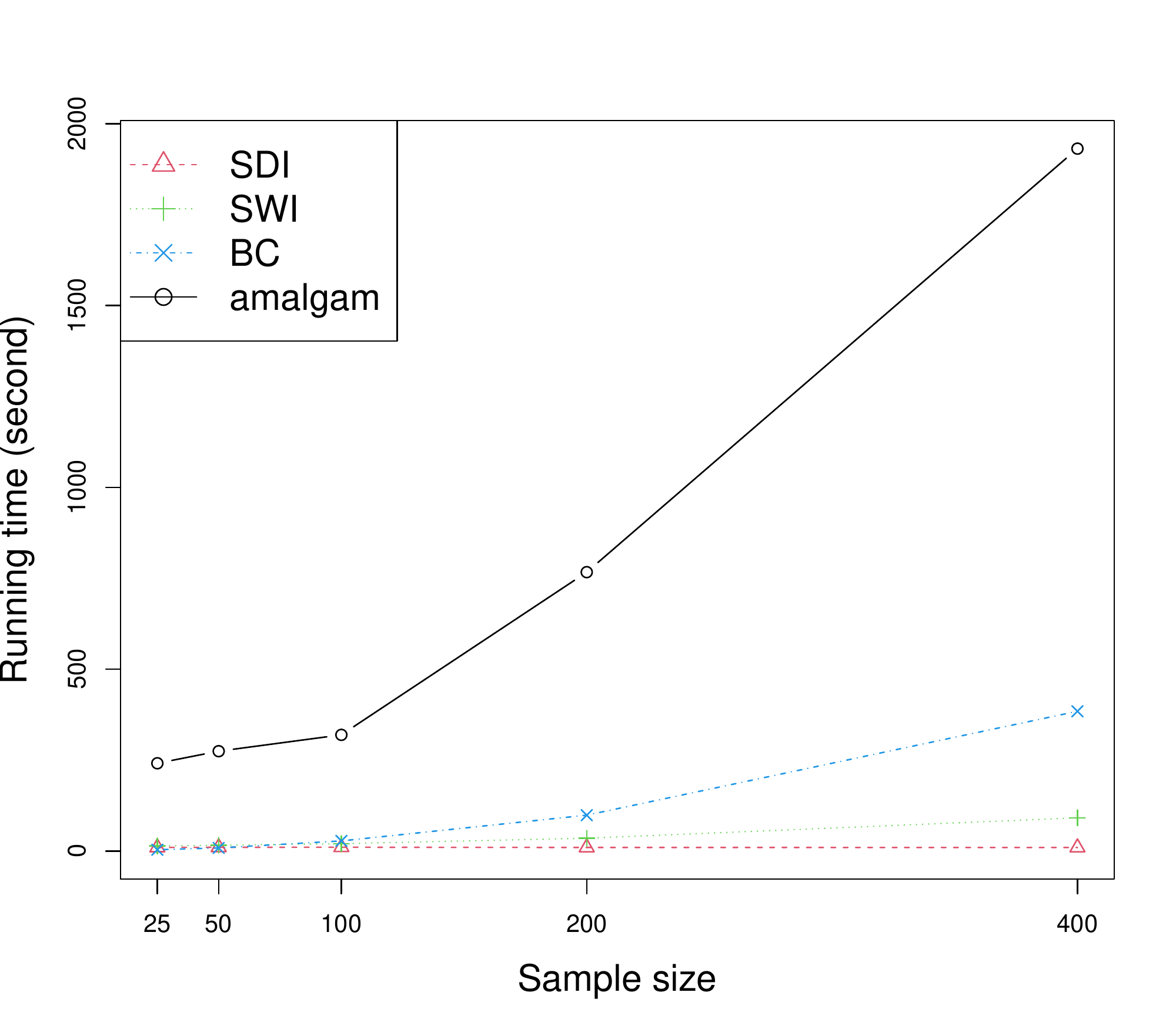}}
  \caption{Simulation: Average running time (in second) of HPAA
    methods with Simpson’s index (SDI), Shannon’s index (SWI) and
    Bray-Curtis dissimilarity (BC), and the \texttt{amalgam} method by
    Quinn and Erb \cite{Quinn2020Amalgam}.}\label{fig:comp_time}
\end{figure*}

We first compare the computation efficiency of HPAA and
\texttt{amalgam}. The results show that HPAA is very computationally
efficient and scales well with the increase of the dimension or the
sample size. In contrast, \texttt{amalgam} is very computationally
intensive even for moderately large $p$ or $n$, making it unsuitable
for large-scale microbiome studies. To explore the effect of the
dimensionality and the the sample size, we generate compositional data
with $n = 100, p \in \{25, 50, 100, 200, 400, 800\}$ and
$n \in \{25, 50, 100, 200, 400\}, p=400$, respectively, using the
\texttt{amalgam} package. The package uses Poisson distribution with
$\lambda = 100$ to generate raw count at each matrix entry and then
converts the counts into compositional data. For \texttt{amalgam}, the
number of amalgamations is fixed at $k=3$ as the computation time of
\texttt{amalgam} greatly increases with the number of amalgamations,
while we use the unconstrained HPAA methods with different loss
functions to generate entire paths of amalgamations. We remark that
the unconstrained HPAA is more time consuming than its constrained
counterpart, as at each step the former always needs to solve the
amalgamation problem over a larger active set. Each setting is
repeated 10 times and the average running time of each method is
reported in Figure~\ref{fig:comp_time}. The results show that HPAA is
very computationally efficient and scales well with the increase of
the dimension or the sample size. In contrast, \texttt{amalgam} is
very computationally intensive even for moderately large $p$ or $n$,
making it unsuitable for large-scale microbiome studies.

\begin{figure}[htp]
  \centering
  \includegraphics[width=0.5\textwidth]{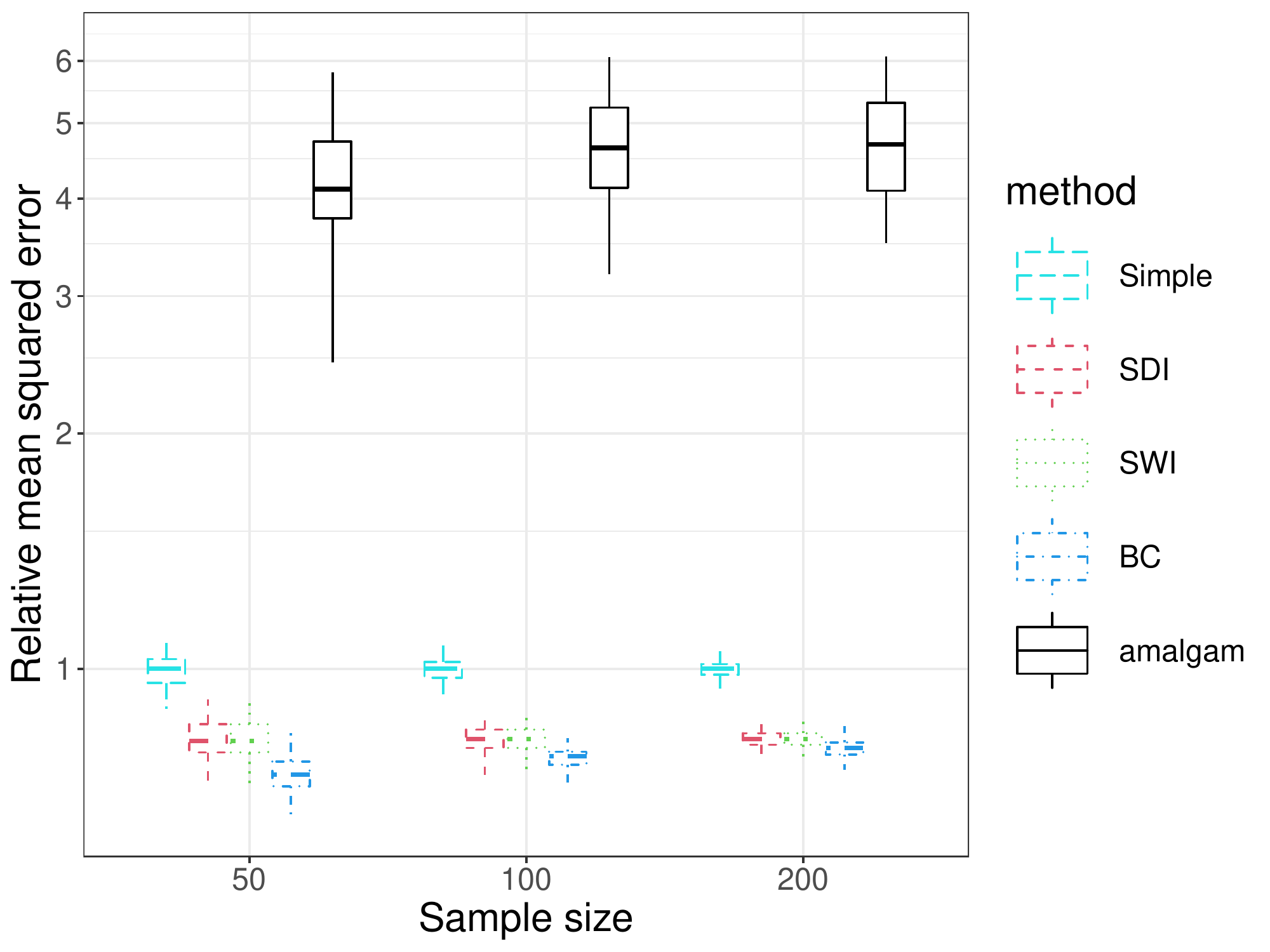}
  \caption{Simulation: Accuracy in preserving between-sample
    Bray-Curtis dissimilarity after dimension reduction. Five
    dimension reduction methods are considered: simple
    prevalence-based filtering method (Simple), HPAA methods with
    Simpson’s index (SDI), Shannon’s index (SWI) and Bray-Curtis
    dissimilarity (BC), and the \texttt{amalgam} method by Quinn and
    Erb \cite{Quinn2020Amalgam}. For each of compression, boxplots are
    constructed for the relative mean squared errors, i.e., mean
    squared error divided by the median of the Simple approach.}
  \label{fig:comp_distance}
\end{figure}

We also compare different dimension reduction methods on how well they
preserve the between-sample distance pattern, which is very importance
in many biological applications. Here we simulate data to mimic the
HIV infection dataset, to be presented in the
Section~4.2 of the main paper. Specifically, each raw
count vector is generated from the multinomial distribution with the
total count being 10,000 and the probabilities being the average
proportions of the $p=60$ taxon in the HIV dataset; the count vector
is then normalized to be compositional. The same taxonomic tree
structure as in the HIV dataset is used. Three sample sizes are
considered, i.e., $n\in \{50,100,200\}$. We use HPAA with weak
taxonomic hierarchy and \texttt{amalgam} to reduce the simulated data
to $k=20$ compositions. The prevalence-based filtering method is also
included as the baseline, which simply keeps the top $20$ most
prevalent taxa. In each simulation, the mean squared error (MSE) of
the two between-sample distance matrices, computed from either the
original data or the reduced data based on Bray-Curtis dissimilarity,
is computed for each method. The procedure is repeated 100 times under
each setting. The results are shown in Figure~\ref{fig:comp_distance},
in which the boxplots are constructed from the relative mean squared
errors (RMSE) using the prevalence-based filtering method as the
baseline, i.e., each RMSE is computed as the MSE divided by the median
of the prevalence-based method (so that the boxplots of the
prevalence-based method are with the median equal to 1). It is clear
that all three HPAA methods with different loss functions outperform
the baseline and the \texttt{amalgam} methods. PAA with the
Bray-Curtis loss performs the best, as it directly aims at preserving
the Bray-Curtis dissimilarity. To our surprise, the \texttt{amalgam}
method performs even worse than the baseline, which may be due to its
requirement of zero-replacement and log-ratio transformation and the
slow convergence of their genetic algorithm.

\clearpage
\bibliographystyle{chicago}
\bibliography{biblio_tensor}

\end{document}